\documentclass[11pt]{article}
\usepackage{a4}
\usepackage{amsmath}
\usepackage{amssymb}
\usepackage{graphicx}
\usepackage{dcolumn}
\usepackage{bm}
\usepackage{citesort}
\begin{document}
\title{How homogeneous are the trehalose, maltose and sucrose water solutions ?
An Insight from Molecular
 Dynamics simulations}
\author{A. Lerbret, P. Bordat, F. Affouard, M. Descamps and F. Migliardo\\
Laboratoire de Dynamique et Structure des Mat\'eriaux Mol\'eculaires\\
UMR CNRS 8024, Universit\'e Lille I, 59655 Villeneuve d'Ascq Cedex France}
\date{}
\maketitle

\begin{abstract}
The structural properties resulting from the reciprocal influence between water and three
well-known homologous disaccharides, namely trehalose, maltose and sucrose, in aqueous solutions have been
investigated in the 4~-~66 wt~\% concentration range by means of molecular dynamics computer simulations. 
Hydration numbers clearly show
that trehalose binds to a larger number of water molecules than do maltose or sucrose, thus affecting the water structure
to a deeper extent. 
Two-dimensional radial distribution functions of trehalose solutions definitely reveal that water is preferentially localized
at the hydration sites found in the trehalose dihydrate crystal, this tendency being enhanced when increasing trehalose concentration.
In a rather wide concentration range (4-49~wt~\%), the fluctuations of the radius of gyration
and of the glycosidic dihedral angles of trehalose indicate a higher
flexibility with respect to maltose and sucrose. 
At sugar concentrations between 33~wt~\% and 66~wt~\%, the mean sugar cluster size and the number of
sugar-sugar hydrogen bonds (HBs) formed within sugar clusters reveal that trehalose
is able to form larger clusters than sucrose but smaller than maltose.
These features suggest that trehalose-water mixtures would
be more homogeneous than the two others, thus reducing both desiccation stresses and ice formation.
\end{abstract}

\section{Introduction}

Disaccharides have received a huge interest in the last decades for their preservation
capabilities of biosystems such as cells, vaccines or therapeutic proteins employed in the food, pharmaceutical
or cosmetics industry~\cite{Franks1985,Wang1999,Haydon2000,Crowe2001}.
Indeed, disaccharides such as trehalose [$\alpha$-D-glucopyranosyl-$\alpha$-D-glucopyranoside], maltose 
[4-O-($\alpha$-D-glucopyranosyl)-$\beta$-D-glucopyranoside] or sucrose
[$\beta$-D-fructofuranosyl-$\alpha$-D-glucopyranoside]
can be added to biologically active solutions to overcome the limited stability range of proteins
 (in pH, in
temperature, in salt concentration, etc.). These additives prevent the partial or even total
degradation of biomolecules due
to the lethal thermal or dehydration stresses encountered
during industrial conservation methods (lyophilization), where trehalose has been found the most
effective~\cite{Crowe1983}. 
Trehalose is also found in high concentration in
organisms which  
enter into a state, called \emph{anhydrobiosis}, where
almost any biological activity is suspended, and thus are able 
to survive conditions of extremely low water content, high or low temperatures.

However, the molecular mechanisms at the origin of the superior capabilities of trehalose and, more generally, of
the biopreservation phenomenon itself still remain unclear, despite various experimental and
theoretical works. Several hypotheses have been proposed but none of them can be considered as fully accepted
since they have sometimes led to contradictory conclusions.

Green and Angell~\cite{Green1989} suggested that dehydrated sugar solutions convert to a glassy
state~\cite{Ediger1996} which has been pictured as acting like amber, encaging molecules and membranes in the same way
that amber traps insects. The higher glass transition temperature $T_{g}$ of trehalose
($T_{g} \approx 393~K$~\cite{Taylor1998,Miller2000}) could explain its greater preservation efficiency compared to maltose
($T_{g} \approx 373~K$~\cite{Taylor1998,Miller2000}) or sucrose ($T_{g} \approx 348~K$~\cite{Taylor1998,Miller2000}) or
other protectants (glucose, sorbitol, xylitol, ...)~\cite{Orford1990,Miller2000,Talja2001}. Nevertheless, it is now well accepted that
biopreservative efficiency does not necessarily scale with the glass transition temperature, as was shown for glycerol~\cite{Caliskan2004},
trehalose-glycerol mixtures~\cite{Cicerone2004} or poly(vinylpyrrolidone) PVP and dextran polymers~\cite{Wang1999,Allison1999}.
Thus, other glass properties such as fragility or anharmonicity or 
intimate interactions of cosolvents with biomolecules via hydrogen bonds (HBs) must be involved.

Crowe \emph{et al.}~\cite{Carpenter1989} suggested that sugar molecules were able to directly interact with the 
polar groups of membranes~\cite{Crowe1984b} or proteins~\cite{Carpenter1989} via HBs, by substituting to the hydration water shell, essential for
proteins structure, dynamics and activity. This would preserve the three-dimensional
structure of biomolecules even at low water content and is referred to as
the $"$Water Replacement$"$ hypothesis~\cite{Crowe1984b}. Trehalose would be able to replace 
more water molecules than other sugars and even to bind to non-polar groups of proteins owing to its higher 
flexibility as proposed by Oku~\emph{et al.}~\cite{Oku2003}.
The water replacement hypothesis seems reasonable for water binding sites
at the protein surface and, indeed, it has been demonstrated from myriads of experiments~\cite{Carpenter1989,Allison1999} that
trehalose can be directly H-bonded to proteins. However, it is less plausible
for internal water molecules. Indeed, it seems improbable that one trehalose molecule could enter
into confined regions of proteins~\cite{Sastry1997}. 

Moreover, Timasheff \emph{et al.}~\cite{Timasheff2002} have shown that many cosolvents
are \emph{preferentially excluded} from the first hydration shell of proteins, at moderate concentrations.
Among these osmolytes, trehalose was found to be
the most preferentially excluded~\cite{Xie1997}, thus inducing the greatest thermodynamic stabilization.
Although this mechanism might not hold at low water contents, Belton and Gil~\cite{Belton1994},
in view of Raman results, proposed that the protein hydration layer is preserved, even at very high dehydration levels.
These water molecules trapped in the glassy matrix formed by trehalose would allow a protein
to preserve its native structure and its internal dynamics necessary to accomplish its
biological function.

Alternatively, an approach based on the \emph{destructuring effect}
of sugars on the water hydrogen bond network (HBN) has been proposed by Magaz\`u\emph{et al.}~\cite{Branca1999b} from numerous experiments such as
infrared and Raman spectroscopies~\cite{Branca1999c,Branca1999b},
nuclear magnetic resonance (NMR) and ultrasound measurements~\cite{Branca1999c}.
They particularly demonstrated that trehalose promotes a more extended hydration
than other disaccharides and binds more strongly to water molecules, thus
preventing more efficiently the crystallization of ice, which causes lethal damages to biosystems.
This hypothesis seems well suited to explain the enhanced cryoprotective efficiency of trehalose, but not its
lyoprotective one.

Finally, Ces\`aro \emph{et al.}~\cite{Cesaro2001} point out the possible role of the polymorphic forms of sugars in the anhydrobiosis
phenomenon. The metastable $\alpha$ form of trehalose, T$_{\alpha}$, is actually produced from the dihydrate crystal 
$\mathrm{T_{2H_{2}O}}$~\cite{Taga1972} only below a given threshold rate of water removal above which dehydrated regions become amorphous~\cite{Willart2003}.
Depending on humidity levels, T$_{\alpha}$ may reversibly transform into $\mathrm{T_{2H_{2}O}}$ within time-scales compatible with the
anhydrobiotic protection \emph{i.~e.} that does not induce rapid changes in volume or internal pressure of cells.
This dehydration-hydration mechanism may play a key role in anhydrobiosis because crystallization of trehalose is easier in Na-Cl salt
aqueous solutions~\cite{Miller1999} and thus in cells owing to the probable presence of charged solutes or macromolecules.

A considerable amount of simulation results has been accumulated on binary sugar/water
solutions~\cite{Brady1993,Liu1997,Roberts1999,Sakurai1997,Bonanno1998,Naidoo2001,Conrad1999,Conrad2001,Engelsen2001,Feick2003,Molinero2003}
or ternary protein or membrane/sugar/water systems~\cite{Sum2003,Pereira2004,Lins2004,Cottone2002}.
Most of studies of binary mixtures deal with the dilute case
(a few wt~\%)~\cite{Brady1993,Liu1997,Sakurai1997,Bonanno1998,Naidoo2001,Engelsen2001} and analyze at ambient temperature ($\approx$ 300~K)
\emph{e.g.} sugar hydration
properties - via radial and orientational distribution
functions, hydration numbers, intrinsic conformations - through adiabatic maps, radius of gyration,
 autocorrelation functions of glycosidic dihedral angles -
and sugar and water mobilities - by means of translational and rotational diffusion coefficients.
Investigations of concentrated mixtures are much fewer.
Roberts~\emph{et al.}~\cite{Roberts1999} studied
 4~-~80~wt~\% aqueous solutions of three homologuous monosaccharides
(namely $\beta$-D-glucose, $\beta$-D-mannose and D-fructose) and pointed out significant effects of the sugar
stereochemistry on the diffusion coefficient of water and
on the structure and relaxation of the HBN in these solutions. Moreover,
Ekdawi-Sever~\emph{et al.}~\cite{Conrad2001,Feick2003}
and Conrad~\emph{et al.}~\cite{Conrad1999} have investigated sucrose and trehalose aqueous solutions from 6~wt~\%
up to 80~wt~\% and even above. They found a larger hydration
number for trehalose~\cite{Conrad2001} for each concentration investigated, as well as
a diffusion coefficient of sucrose consistently larger than that of trehalose at high
concentration (72 wt~\% or above)~\cite{Feick2003}. Finally, Molinero ~\emph{et al.}~\cite{Molinero2003}
studied concentrated water-sucrose systems (50-100 wt~\%). They showed percolation of the sucrose HBN in the range of
60~-~67~wt~\% at 333~K and suggested that the formation of sucrose network may increase both the mixture resistance to shear
deformation and mechanical stability observed experimentally.

Simulations of ternary systems have also provided relevant results.
Sum~\emph{et al.}~\cite{Sum2003} and Pereira~\emph{et al.}~\cite{Pereira2004} published results on simulations of lipid membrane in
presence of sugars, and found evidence for direct HBs with specific parts of the lipid molecules, consistently with previous experiments.
Moreover, Lins~\emph{et al.}~\cite{Lins2004}
proposed a model of trehalose-lysozyme interaction on the nanosecond scale and at moderate concentration (18 wt~\%) in
agreement with the suggestion of Belton and Gil~\cite{Belton1994}, which was also inferred by Cottone~\emph{et al.}~\cite{Cottone2002} from
simulations of carboxy-myoglobin in trehalose (50 wt~\% - 89 wt~\%).

In order to get a deeper understanding of disaccharide-water
solutions structural properties in the framework of the biopreservation problem, we have carried out a careful comparative
study of trehalose, maltose and sucrose in aqueous
solutions  
by molecular dynamics (MD) simulations.
The H-bonding capabilities of these three sugars
are directly comparable since they possess
the same chemical formula $\mathrm{C_{12}H_{22}O_{11}}$ and the same number of OH groups.
Water-water, water-sugar and both intra and inter sugar-sugar properties and their relationships have been probed
in order to measure the homogeneity of the different mixtures. Different relevant structural parameters have been used: 
partial static structure factors, two-dimensional (2D) radial distribution functions, probability of HB formation, water clusters size, hydration number, 
molecular flexibility and sugar clusters size. 
The main goal of this work was to
analyze  the physical properties changes of the water/sugar solutions upon increasing the sugar concentration,
to identify pertinent parameters aiming to explain
the superior bioprotective effectiveness of trehalose and 
to discuss these results in the framework of the different suggested hypotheses. A schematic model 
of the structure of sugar-water mixtures showing the destructuring effects of the different sugars on the water HBN is proposed.

\section{Details of the simulation}
\begin{figure}[h]
\includegraphics[width=8.2cm,clip=true]{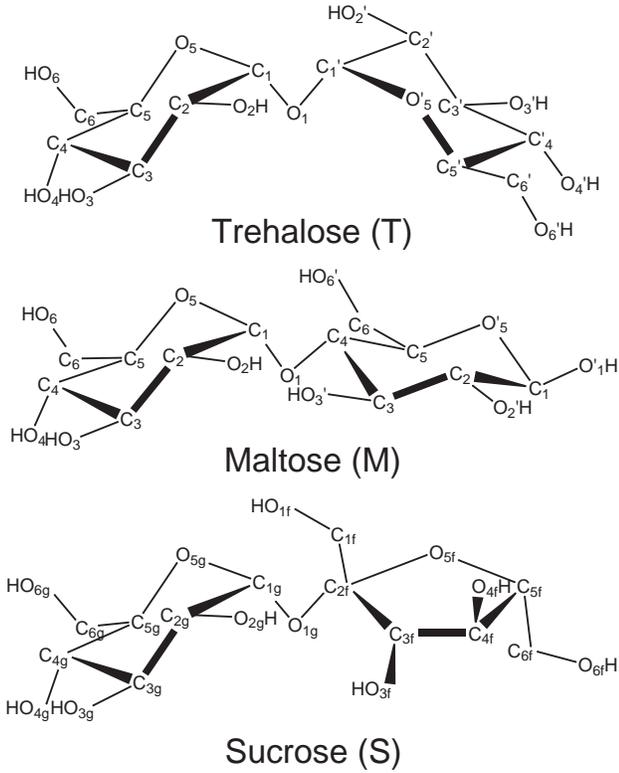}
\caption{\label{fig0} Schematic representation of the studied disaccharides : trehalose (T), maltose (M), and sucrose (S).
Only hydrogen atoms belonging to hydroxyl groups have been represented for clarity reasons. Moreover, the glucose and fructose rings of sucrose are
identified by the \emph{g} and \emph{f} subscripts, respectively.
}

\end{figure}

MD simulations have been performed using the molecular dynamics
package DL\_POLY\_2 \cite{DLPOLY}.
$\alpha,\alpha$-trehalose,
$\beta$-maltose and sucrose disaccharides (see Fig.~\ref{fig0}) have been investigated
and are referred to in the remaining part of the paper as trehalose (T),
maltose (M) and sucrose (S), respectively.
Disaccharide molecules have been considered flexible and have been
modeled using the well-known all-atom force field designed
for carbohydrates developed by Ha \emph{et al.}~\cite{Ha1988} which
has been extensively employed~\cite{Brady1993,Engelsen1995,Liu1997}.
The SPC/E model~\cite{Berendsen1987} has been used to represent water molecules whose
geometry has been constrained using the SHAKE algorithm~\cite{Rickaert1977}.
This model is known to give a too stiffly structured water but
it realistically describes the diffusion coefficient of water at ambient temperature.
In order to check the dependence of our results on the water model,
some additional simulations have been performed using the TIP3P water model~\cite{Jorgensen1983}.
A cutoff radius of 10 \AA \ has been used to account for non-bonded interactions.
Electrostatic interactions have been handled by the reaction-field method using the permittivity constant $\epsilon$ = 72.
Since a concentrated disaccharide solution has a relatively low dielectric constant, we checked the validity of the 
reaction-field method by performing one MD simulation of the 66~wt~\% sucrose solution at $T$ = 373 K
using an Ewald summation. No significant structural or dynamical difference has been found between both techniques.
The compressibility of the simulated solutions was that of water (it is a parameter of the DL\_POLY\_2 program).
A Lennard-Jones potential has been employed to represent van der Waals interactions
and Lorentz-Berthelot mixing-rules have been used for cross-interaction terms.
MD simulations have been realized in the NPT statistical ensemble where the number of molecules $N$,
pressure $P$ and temperature $T$ are fixed. The pressure has been set to 1.0~bar using weak
coupling to a pressure bath (Berendsen barostat~\cite{Berendsen1984}) with a relaxation time of 1.0 ps. The
investigated temperatures range from 273 K up to 373 K in steps of 20 K. They have been
maintained constant during a given simulation using weak coupling to a heat bath (Berendsen thermostat~\cite{Berendsen1984})
with a relaxation time of 0.1 ps.
Newton's equation of motions have been integrated using the Verlet leapfrog algorithm~\cite{Allen1989}.
The simulation lengths range from 200 ps to 2 ns depending on concentrations and
temperatures with a time step of 0.5 fs for the binary aqueous solutions and 2 fs for
pure water. The 0.5-fs timestep is justified by the flexible O-H bonds of the
carbohydrates hydroxyl groups.
The disaccharide initial conformations have been deduced from neutron
and X-ray studies (trehalose~\cite{Taga1972}, maltose~\cite{Gress1977} and sucrose~\cite{Brown1973}).
Starting positions of the sugar and the water molecules have been arbitrary chosen on a cubic lattice.
In order to remove local contacts that result from this construction, short simulations with a
thermostated temperature of 0 K and a small timestep have been performed.
Simulated systems are composed of $N_{W}=512 $ water molecules and $N_{S}$~=~0, 1, 5, 13, 26 or 52 sugars
molecules (either trehalose, maltose or sucrose) corresponding to weight concentrations
of 0, 4, 16, 33, 49 or 66~\%, respectively. These systems have been considered to be large
enough to give size-independent results. Moreover, they cover a relatively broad concentration
range.
Depending on the temperature and concentration, each system has been equilibrated
from 50 ps up to 500 ps, and configurations have then been saved with time intervals that ranged
between 0.05 ps to 0.5 ps. Table~\ref{table1} summarizes simulation data for $T$ = 293 K.

\begin{table*}[htbp]
\centering
\caption{\label {table1}
System compositions, densities, and equilibration/simulation times at $T$ = 293 K at 
different sugar concentrations $\phi$. Data corresponding to $\phi$ = 0~wt~\% result
from only one simulation of pure water.
}
\vspace* {1.0cm}
\begin{tabular}{cccccccc}
\cline {1-5}
\hline
$\phi$ (wt \%) & no. sugars/no. H$_{2}$O & \multicolumn{3}{c}{density ($g.cm^{-3}$)} & \multicolumn{3}{c}{Eq./Sim. time (ns) }\\
\cline {3-8}
  &  &\multicolumn{1}{c}{T}&\multicolumn{1}{c}{M}&\multicolumn{1}{c}{S}&\multicolumn{1}{c}{T}&\multicolumn{1}{c}{M}&\multicolumn{1}{c}{S}\\
\cline {1-8}
0 & 0/512 & \multicolumn{1}{c}{1.005} & \multicolumn{1}{c}{1.005} & \multicolumn{1}{c}{1.005} & \multicolumn{1}{c}{0.1/0.3} & \multicolumn{1}{c}{0.1/0.3}& \multicolumn{1}{c}{0.1/0.3}\\
4 & 1/512 & \multicolumn{1}{c}{1.020} &\multicolumn{1}{c}{1.020} &\multicolumn{1}{c}{1.020}&\multicolumn{1}{c}{0.1/0.4} &\multicolumn{1}{c}{0.1/0.4} &\multicolumn{1}{c}{0.1/0.4}\\
16 & 5/512 & \multicolumn{1}{c}{1.076} &\multicolumn{1}{c}{1.078} &\multicolumn{1}{c}{1.076}& \multicolumn{1}{c}{0.1/0.4} &\multicolumn{1}{c}{0.1/0.2} &\multicolumn{1}{c}{0.05/0.15}\\
33 & 13/512 & \multicolumn{1}{c}{1.167} &\multicolumn{1}{c}{1.168} &\multicolumn{1}{c}{1.168}& \multicolumn{1}{c}{0.1/0.4} &\multicolumn{1}{c}{0.15/0.35} &\multicolumn{1}{c}{0.125/0.375}\\
49 & 26/512 & \multicolumn{1}{c}{1.269} &\multicolumn{1}{c}{1.276} &\multicolumn{1}{c}{1.273}& \multicolumn{1}{c}{0.2/0.8} &\multicolumn{1}{c}{0.3/0.7} &\multicolumn{1}{c}{0.125/0.375}\\
66 & 52/512 & \multicolumn{1}{c}{1.379} &\multicolumn{1}{c}{1.392} &\multicolumn{1}{c}{1.377} & \multicolumn{1}{c}{0.25/1.75} &\multicolumn{1}{c}{0.5/1.0} &\multicolumn{1}{c}{0.25/1.75}\\
                                                                                                                                             
\end{tabular}
\end{table*}

\section{Water-sugar interactions}

\subsection{Comparison of experimental and simulated static structure factors}
As a means to get some insights on the structure of disaccharide-water solutions, we have computed the static structure factor S(\textbf{Q}), which
can be classically obtained from coherent neutron scattering experiments~\cite{Dolling1979} from $S(\bf{Q})=<|\rho_{\textbf{Q}}|^2>$, where
$\rho_{\textbf{Q}}$ is the time-dependent density correlator defined as:
$\rho_{\textbf{Q}}(t)=\Sigma{b_{\alpha}.exp[\textbf{\emph{i}}.\textbf{Q}.\textbf{r}_{\alpha}]}$. The sum is over all the $\alpha$ atoms
of the system, $b_{\alpha}$ and $\textbf{r}_{\alpha}$ are the coherent scattering length and the position of the $\alpha$ atom, respectively.
An average over isotropically distributed $\textbf{Q}$-vectors which have the same modulus Q is performed in order to get S(Q). Partial
static structure factors $S_{HH}(Q)$ and $S_{XX}(Q)$ are obtained by only including or only excluding
in the summation both the hydrogen atoms of sugar and water molecules.

\begin{figure}[h]
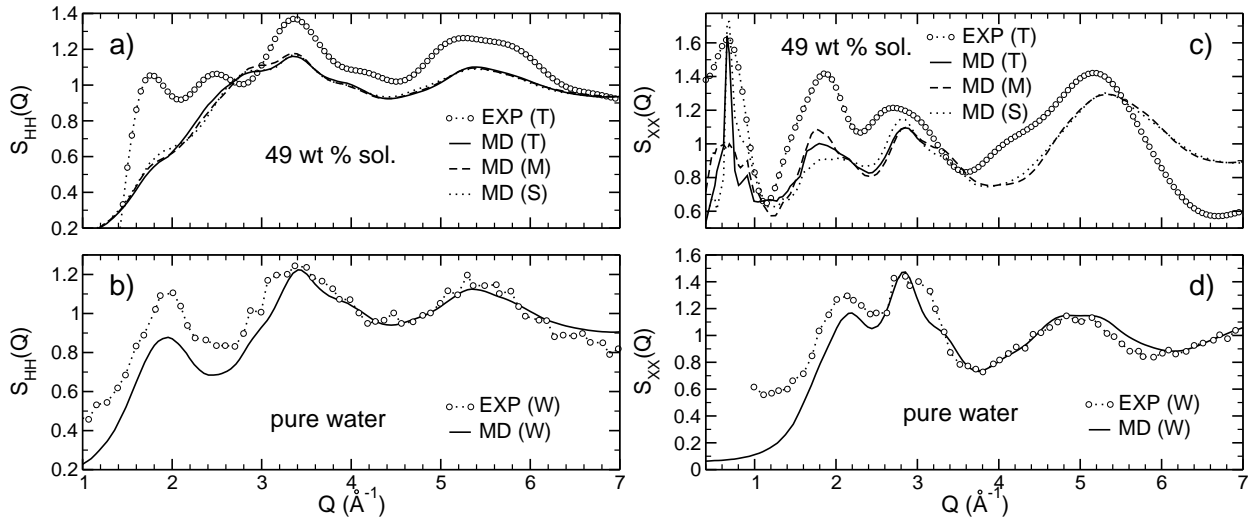

\includegraphics[width=8.2cm,clip=true]{figure2.eps}
\includegraphics[width=8.2cm,clip=true]{figure2_bis.eps}
\caption{\label{fig0.1}
Partial static structure factors $S_{HH}(Q)$ of hydrogen atoms (a) and $S_{XX}(Q)$ of all atoms distinct
from hydrogen (c) for the 49 wt~\% sugar-water solutions obtained from the present MD investigation
at $T$ = 293 K and from experiments (for trehalose only) at $T$ = 300 K~\cite{Cesaro2004,Branca2002bis}.
Data for pure water obtained from MD simulation and experiment~\cite{Soper1986} are also included ((b) and (d)).
}
\end{figure}

Figure~\ref{fig0.1} shows a comparison between the partial static structure factors $S_{HH}(Q)$ and $S_{XX}(Q)$
of the 49~wt\% trehalose-water solutions and of pure
water obtained experimentally from neutron diffraction~\cite{Branca2002bis,Cesaro2004} and those from the present MD study. 
The $S_{HH}(Q)$ and $S_{XX}(Q)$ of the 49~wt\%
sucrose and maltose aqueous solutions are also given for comparison, although no experimental data is available.
The agreement found between both techniques is good and found better for the $S_{XX}(Q)$ than for the $S_{HH}(Q)$, particularly at Q 
$\leqslant$ 3 \AA$^{-1}$. This arises from the difficulty to reproduce precisely relative positions of hydrogen atoms, which strongly
depend on the orientation of water molecules and of sugar hydroxyl groups. The peak observed at Q $\approx$ 2 \AA$^{-1}$ for pure water seems to decrease
with the addition of sugars. This could be representative of a lowered tetrahedral order of water molecules. Nevertheless, 
no striking and meaningful difference is observed between the partial $S_{HH}(Q)$ of the three sugar-water solutions.
Similarly, the modification of the peak observed at Q  $\approx$ 2.2 \AA$^{-1}$ in the $S_{XX}(Q)$ of pure water, representative of 
H-bonded water molecules, may suggest a 
decrease of the tetrahedrality of water upon increasing the sugar concentration.
However, the rather large contributions from sugar molecules in the $S_{HH}(Q)$ and $S_{XX}(Q)$ of mixed solutions (about one third and 
half of the number of considered atoms for the calculations, respectively) makes this kind of analysis quiet hazardous.
Therefore, the shift
observed in the $S_{XX}(Q)$ of the peak at about Q $\approx$ 2 \AA$^{-1}$ for pure water may be attributed 
to differences in the conformation and topology of sugars.
Interestingly, significant discrepancies are observed between the three studied solutions at Q $\leqslant$ 1 \AA$^{-1}$, which represents
distances of intermolecular interactions between sugars, and where the water contribution is quiet low.
This motivates a more detailed analysis, which could be performed by separating 
contributions from both sugar and from water molecules, and therefore probe their complex structures.
However, a careful analysis of the HBN of these solutions will be presented in a different manner in the following.

\subsection{Crystallinity of trehalose/water solutions}
As indicated in the introduction, some authors suggested that the rich crystalline polymorphism of trehalose
could be involved in the origin of its superior bioprotective capability~\cite{Cesaro2001}.
From neutron scattering experiments, Magaz\`u~\emph{et al.} recently showed that trehalose possesses a more crystalline character 
than sucrose and maltose which could provide a more rigid environment 
for protecting biological structures~\cite{Magazu2004}. Furthermore, 
Engelsen~\emph{et al.}~\cite{Engelsen2000}  demonstrated that 
the trehalose hydration pattern in dilute solution ($\approx$ 4 wt\%) 
resembles that of the trehalose dihydrate crystal~\cite{Taga1972}.
Inspired by this latter work, we compared the hydration sites 
of high concentrated water-trehalose solutions
with those of the trehalose dihydrate crystalline phase.
2D radial $g(r_{1},r_{2})$ distributions functions have been computed and are 
defined as:

\begin{equation}
g(r_{1},r_{2}) = \frac{N(r_{1},r_{2})}{\rho_{W}.V_{intersect}(r_{1},r_{2},\Delta r)}\hbox{\hspace{0.1cm}} \label{gr1r2}
\end{equation}

\begin{figure}[h]
\includegraphics[width=8.2cm,clip=true]{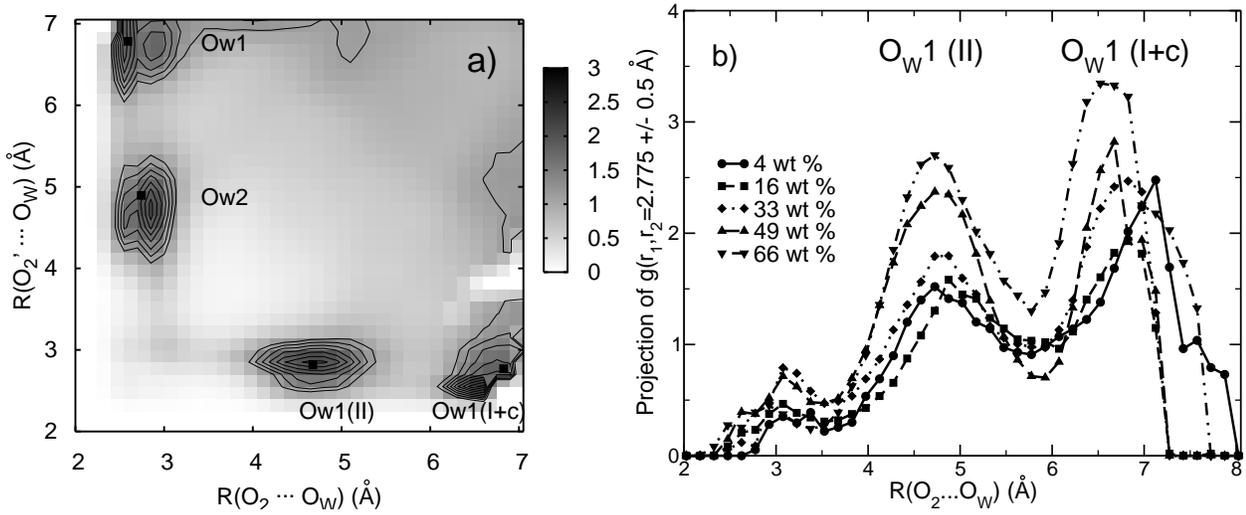}
\includegraphics[width=8.2cm,clip=true]{figure3_bis.eps}
\caption{\label{fig0.2}
(a) Normalized $g(r_{1},r_{2})$ pair distribution of the 49 wt~\% trehalose-water solution at 293 K. $r_{1}$ and $r_{2}$
correspond to the distance between the water oxygen atoms and the  $\mathrm{O_{2}}$ and $\mathrm{O'_{2}}$ oxygen atoms
of the trehalose molecules, respectively.
Black squares indicate the position of the two water molecules of the trehalose dihydrate crystal, $O_{W}1$ and $O_{W}2$. Symmetries are indicated
in parentheses (see text and ref.~\cite{Engelsen2000}).
(b) Projection of $g(r_{1},r_{2})$ at $r_{2}$ = 2.78 $\pm$ 0.5 \AA~for the different trehalose-water solutions at 293 K.}
\end{figure}

The exhaustive description of $g(r_{1},r_{2})$ functions is described in ref.~\cite{Engelsen2000},
so we give here only the essential details.
 $g(r_{1},r_{2})$ gives the probability of finding a water oxygen atom
at a distance $r_{1}$ and $r_{2}$ from two selected solute oxygen atoms, $\mathrm{O_{2}}$ and $\mathrm{O'_{2}}$ (see Fig.~\ref{fig0}),
relative to the probability expected for a random distribution.
$N(r_{1},r_{2})$ denotes the number of water oxygens at distances $r_{1}$ and $r_{2}$ 
from the two selected solute atoms, averaged over
all the solute molecules and the configurations of the simulation. 
$\rho_{W}=N_{W}/V$ stands for the mean water density of the simulated system
and $V_{intersect}(r_{1},r_{2},\Delta r)$ designates the intersection volume between the two shells of inner radii $r_{1}$ and $r_{2}$, 
and thickness $\Delta r$. 
Figure~\ref{fig0.2} (a) shows the normalized $g(r_{1},r_{2})$ pair distribution of 
the 49 wt~\% trehalose-water solution at 293 K. It is clearly seen that water molecules are localized in particular regions of 
the 2D distribution, the maximum of which corresponds to the hydration sites
found in the dihydrate crystal. Four sites are particularly observed: $\mathrm{O_{W}1}$, $\mathrm{O_{W}2}$, $\mathrm{O_{W}1(II)}$
 and $\mathrm{O_{W}1(I+c)}$ following the same nomenclature as in
ref.~\cite{Engelsen2000}) where I~and II denotes the trehalose dihydrate $\mathrm{P2_{1}2_{1}2_{1}}$ symmetry operations 
 $(x,y,z)$ and $(-x+\frac{1}{2},-y,z)$, respectively.
A good agreement has been found
with the $g(r_{1},r_{2})$ of ref.~\cite{Engelsen2000} for the 4~wt~\% solutions, with both the TIP3P and SPC/E water models (data not shown).
The symmetry observed in the hydration pattern of trehalose is consistent with the approximate twofold symmetry found in the trehalose
dihydrate crystal~\cite{Taga1972}. This means that the most probable trehalose conformation in solution is rather close to the conformation 
in the dihydrate crystal, as shown in ref.~\cite{Liu1997,Engelsen2000}.
The concentration dependence of the water density at 293 K is depicted in 
Fig.~\ref{fig0.2} (b) which shows
the $g(r_{1},r_{2})$ 
at $r_{2}$ = 2.78 $\pm$ 0.5 \AA~\emph{i.e.} in the region where the $O_{W}1$ water molecule of the dihydrate structure should be found. 

A significant increase of the localization
of water molecules at the $\mathrm{O_{W}1}$ and $\mathrm{O_{W}2}$ (not shown) hydration sites of the trehalose dihydrate crystal is observed. 
This result reveals that the position of the water molecules is more and more sterically restricted as the concentration of sugar increases.
It clearly supports the results obtained by Magaz\`u~\emph{et al.} on the 
crystallinity of trehalose/water solutions~\cite{Magazu2004}.
It should be noted that similar results were obtained for maltose/water solution
for which water molecules are found to be preferentially 
localized on the water site of the monohydrate structure (data not shown)~\cite{Gress1977}.
The localization of water molecules seen in the present investigation is consistent
with the suggestion of Aldous~\emph{et al.}~\cite{Aldous1995} that dihydrate trehalose state works as a sponge 
for water molecules.
Indeed, the air-exposed part of the glassy trehalose may quiet easily transform into the dihydrate crystal form which could further 
stabilize the sugar matrix. On the contrary, a small amount of absorbed water will readily plasticize the sucrose matrix, since it cannot form 
any hydrated crystalline phase. Finally, maltose should be intermediate between trehalose and sucrose, since it may exist in a monohydrate
crystalline form. Hydration properties of sugars will be analyzed in the following.

\subsection{Hydration numbers}
Since trehalose, maltose and sucrose possess the same number of hydroxyl groups,
the average number of water molecules H-bonded per disaccharide, provides
a useful parameter to quantify the water-disaccharide interaction. This parameter 
will be identified as the hydration number $n_H$
in the following although it often refers in the literature to the total number of water 
in the solvation sphere~\cite{Engelsen2000}.

Two different geometric criteria commonly used in molecular dynamics studies~\cite{Roberts1999,Liu1997}
were defined and called I and II in the following. In the present study, two molecules
are considered to be H-bonded if the oxygen-oxygen distance $d_{OO}$ is less than
3.4 \AA \ and the O-H$\cdot\cdot\cdot$O angle larger than 160 deg. or 120 deg. for the criterion I or II,
respectively. The criterion I~\cite{Roberts1999} relates to relatively well-formed and thus
strong HBs
whereas the less stringent criterion II~\cite{Liu1997} also includes more deformed, weaker HBs. Owing to the strong dependence of the HB
energy on the O-H$\cdot\cdot\cdot$O angle, both criteria may probe different
populations of HBs and will therefore be used in the present paper. Moreover, we have not considered energetic criteria because
of the difficulty to define with any ambiguity water-solute HBs, as pointed out by Brady~\emph{et al.}~\cite{Brady1993}.

\begin{figure}[h]
\includegraphics[width=8.2cm,clip=true]{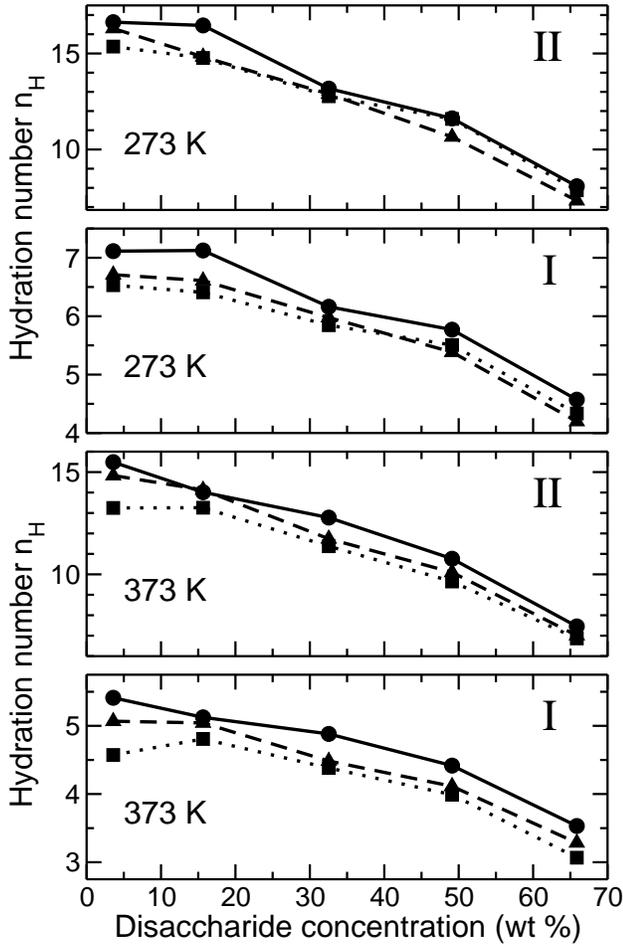}
\caption{\label{fig1}
Hydration number $n_H$ of trehalose (solid line), maltose (dashed line) and sucrose (dotted line)
as a function of their concentration at two temperatures (273 K and 373 K) and for both geometric criteria I and II
(see text for definition of geometric criteria).}
\end{figure}

Figure~\ref{fig1} shows the hydration number $n_H$ of the three studied disaccharides
as a function of the disaccharide concentration at $T$ = 273 K and 373 K.
We clearly see that $n_H$ decreases monotonically by a factor of about 2 for both criteria when the concentration increases from
4 wt~\% to 66 wt~\%.
For the most dilute solutions, disaccharides are surrounded by several hydration layers and
their structure mainly results from their intrinsic interaction with water.
At higher concentrations, disaccharides begin to share water molecules
and to form sugar-sugar HBs. Table~\ref{table3} presents the sugar-sugar HBs statistics (criterion II) at $T$ = 293 K.
As expected, the normalized mean number of intermolecular sugar-sugar HBs steadily increases
with concentration. But most important, significant differences
appear among sugars. Indeed, sucrose (maltose) forms the lowest (largest) 
number of HBs in the concentration range where inter-HBs become
meaningful \emph{i.~e.} 33~-~66 wt~\%. Trehalose forms slightly less inter-HBs than maltose does. This has direct consequences in the
clustering differences between sugars, as will be shown in section~\ref{sugar_cluster} of this paper.

\begin{table*}[htbp]
\centering
\caption{\label {table3}
Normalized mean number of sugar-sugar intermolecular $<n_{HB}>_{inter}$/$N_S$ and intramolecular 
$<n_{HB}>_{intra}$/$N_S$ HBs for the sugar-water solutions at $T$ = 293 K
at different sugar concentrations $\phi$ (the brackets $< >$ denote time-averaging over stored configurations
and $N_S$ designates the number of sugars in the simulation box). }
\vspace* {1.0cm}
\begin{tabular}{ccccccc}
\cline {1-5}
\hline
$\phi$ (wt \%) & \multicolumn{3}{c}{$<n_{HB}>_{inter}$/$N_S$} & \multicolumn{3}{c}{$<n_{HB}>_{intra}$/$N_S$}\\
\cline {2-7}
  & \multicolumn{1}{c}{T}&\multicolumn{1}{c}{M}&\multicolumn{1}{c}{S}&\multicolumn{1}{c}{T}&\multicolumn{1}{c}{M}&\multicolumn{1}{c}{S}\\
\cline {1-7}
4 & \multicolumn{1}{c}{0.000} &\multicolumn{1}{c}{0.000} &\multicolumn{1}{c}{0.000}&\multicolumn{1}{c}{0.304} &\multicolumn{1}{c}{0.739} &\multicolumn{1}{c}{2.490}\\
16 &\multicolumn{1}{c}{0.153} &\multicolumn{1}{c}{0.361} &\multicolumn{1}{c}{0.266}& \multicolumn{1}{c}{0.146} &\multicolumn{1}{c}{0.455} &\multicolumn{1}{c}{1.962}\\
33 & \multicolumn{1}{c}{0.767} &\multicolumn{1}{c}{0.853} &\multicolumn{1}{c}{0.435}& \multicolumn{1}{c}{0.141} &\multicolumn{1}{c}{0.373} &\multicolumn{1}{c}{2.058}\\
49 &\multicolumn{1}{c}{1.322} &\multicolumn{1}{c}{1.942} &\multicolumn{1}{c}{0.738}& \multicolumn{1}{c}{0.451} &\multicolumn{1}{c}{0.346} &\multicolumn{1}{c}{2.210}\\
66 & \multicolumn{1}{c}{2.689} &\multicolumn{1}{c}{2.985} &\multicolumn{1}{c}{2.293} & \multicolumn{1}{c}{0.175} &\multicolumn{1}{c}{0.400} &\multicolumn{1}{c}{1.979}\\

\end{tabular}
\end{table*}

From Fig.~\ref{fig1}, trehalose is found to be more hydrated than sucrose and maltose, whatever the concentration
and the geometric criterion considered, in good agreement with Ekdawi-Sever~\emph{et al.}~\cite{Conrad2001}.
They have obtained similar results
when comparing sucrose and trehalose at 353 K and 360 K, respectively,
in the 6~-~90~wt~\% concentration range. They particularly suggested that 
the intramolecular HBs were at the 
origin of the difference observed between both sugars.
Table~\ref{table3} clearly confirms this hypothesis. 
It is indeed well-known that maltose and sucrose form intra-HBs in their
crystalline state on the contrary to trehalose~\cite{Taga1972,Gress1977,Brown1973}. The crystalline $\beta$-maltose
forms one intra-HB between $\mathrm{O_{2}-HO_{2}}$ and $\mathrm{O_{3}'}$. Two intra-HBs,
namely $\mathrm{O_{6f}-HO_{6f}}\cdot\cdot\cdot\mathrm{O_{5g}}$ and $\mathrm{O_{1f}-HO_{1f}}\cdot\cdot\cdot\mathrm{O_{2g}}$ are present in
crystalline sucrose. As a consequence, the disaccharides HB
centers (i.e. hydroxyl hydrogen and oxygen atoms) involved in intra-HBs no
longer remain available for water-disaccharide HBs. The large number of intra-HBs of sucrose relative to trehalose and maltose 
for the 4~wt~\% concentration explains why its hydration number is significantly lower,
while those of maltose and trehalose are rather similar.
Moreover, a crossover between maltose and sucrose hydration curves is found at temperatures
lower than 353 K, as exemplified in Fig.~\ref{fig1} at $T$ = 273 K. This could arise
from different concentration dependences of the relative number of intra-HBs and inter-HBs between sucrose and maltose. Indeed,
the difference between $<n_{HB}>_{inter}$/$N_S$ of maltose and sucrose steeply raises above 33~wt~\% disaccharide 
concentration. Meanwhile, the difference between $<n_{HB}>_{intra}$/$N_S$ of maltose and sucrose changes moderately. Therefore, the overall
effect is a more pronounced decrease of maltose hydration numbers compared to sucrose when increasing disaccharide concentration.

Hydration numbers of the 4 wt~\% disaccharide solutions at $T$ = 293 K are given in Table~\ref{table2}.
Trehalose clearly binds to a larger number of water molecules and consequently the water structure
will be more affected by this sugar than the others as it will be shown in
a next section of this paper devoted to water structure.
The higher hydration number of trehalose is also seen to be independent on the choice
of the model of water as proved by the additional calculations
performed using the TIP3P water model (see Table~\ref{table2}). A good agreement
is particularly obtained between both SPC/E and TIP3P models.
$n_H$ (I) obtained from the SPC/E water model is slightly higher than with TIP3P. This may be understood
as the stronger water-disaccharide HBs formed with SPC/E by means of the larger water partial charges.
In addition, hydration numbers calculated in this study may rather be dependent on the potential energy landscape (PEL)
of sugars explored within the length of simulations. For computation time reasons, it was not possible to explore each
minima of the adiabatic map of sugars. Moreover, only the $\beta$-anomeric form of maltose was considered
here, whereas an equilibrium between the $\beta$- and the $\alpha$-form of maltose establishes in reality.
Several hydration numbers found in the literature from experimental and other numerical studies
are also reported in Table~\ref{table2} for comparison.
A fair agreement is obtained with our present results and confirm
that trehalose always possess the higher hydration number.
Discrepancies can be due to
the very different definitions of the hydration number found in the literature and the
various approaches and models used to derive hydration numbers from ultrasound,
DSC, viscosity, or water activity measurements.
Nevertheless, the two geometric criteria employed in this study seem to be well suited to
distinguish,
within the first hydration shell, strongly H-bonded (I) water molecules from weakly
H-bonded (II) water. These two types of HBs
could be at the origin of the differences observed experimentally.

\begin{table*}[htbp]
\centering
\caption{\label {table2}
Hydration numbers $n_H$ of trehalose, maltose and sucrose molecules computed from the present
investigation using both SPC/E and TIP3P models at $T$ = 293 K. $n_H$ obtained from other computer simulations
($^{*}$, $d_{OO} \leq $ 3.5 \AA \ and $^{**}$, $d_{OO} \leq 2.8 $ \AA)
and experiments (acoustic~\cite{Galema1991,Branca2001}, viscosity~\cite{Branca2001} and
calorimetric~\cite{Kawai1992,Furuki2002} measurements)
are also indicated for comparison.
}
\vspace* {1.0cm}
\begin{tabular}{ccccc}
\hline
$n_H$ & SPC/E & TIP3P & other MDs & Experiments \\
\hline
Trehalose & 6.5 (I) & 6.1 (I) & 7.8$^{*}$\cite{Engelsen2000},13.4\cite{Conrad1999},18.9\cite{Bonanno1998} & 7.95\cite{Kawai1992},10.9\cite{Furuki2002},12.1\cite{Branca2001}  \\
    & 16.4 (II) & 16.8 (II) &  22.5\cite{Liu1997},27.5$^{**}$\cite{Engelsen2000} & 15.2\cite{Branca2001},15.3\cite{Galema1991} \\\\
Maltose & 6.5 (I) & 5.7 (I) & 10-11\cite{Fringant1995}, 22.6\cite{Brady1993} & 6.50\cite{Kawai1992},9.5\cite{Furuki2002},11.7\cite{Branca2001}  \\
  & 16.1 (II)  &  16.1 (II) &   & 14.2\cite{Branca2001},14.5\cite{Galema1991}  \\\\
Sucrose & 6.2 (I) & 5.6 (I) & 7.0$^{*}$\cite{Engelsen1996}, 11.7\cite{Conrad2001}, 24.7$^{**}$\cite{Engelsen1996} & 6.33\cite{Kawai1992}, 8.5\cite{Furuki2002}, 11.2\cite{Branca2001} \\
  & 14.8 (II) & 15.1 (II) & & 13.8\cite{Branca2001}, 13.9\cite{Galema1991}  \\\\
\end{tabular}
\end{table*}

\subsection{HB probability}

The hydrogen bond network of water may be described by the fractions $f_{j}$ of water
molecules that form $j$ HBs with their neighboring water molecules~\cite{Stanley1980}. Indeed, water molecules would
be involved in four water-water linear HBs (two as proton donor and two as proton acceptor) in
perfect ice, because each one is surrounded by four neighbors which form a perfect tetrahedron.
In real water, deviations from this ideal view occur and bifurcated HBs as well as non-H-bonded
hydroxyl groups exist~\cite{Giguere1987}.
Raman scattering studies on the O-H stretching band of water as function of temperature reveal the
existence of an isobestic point~\cite{Darrigo1981}. From this observation, some authors proposed an arbitrary decomposition
of the Raman spectrum of water into two temperature-independent classes of water molecules,
denoted as 'open' (or intact bond) and 'close' (or broken bond). The former was
assigned to tetrahedrally bonded water molecules, while the latter to the remaining water molecules,
including bifurcated HBs, free OH groups, etc.

\begin{figure}
\includegraphics[width=8.2cm,clip=true]{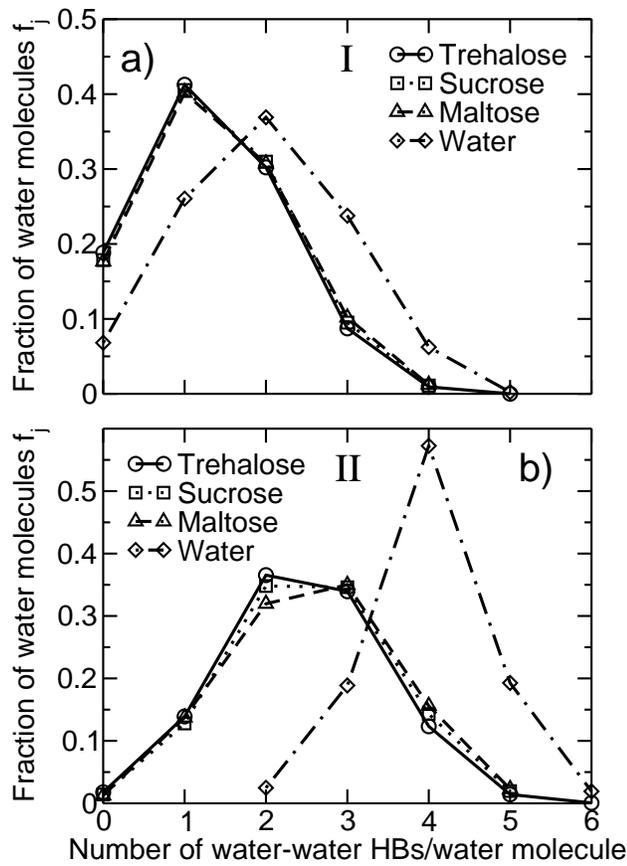}
\caption{\label{fig6}
Fractions $f_{j}$ of water molecules forming $j$ HBs with other neighboring water molecules for the
66~wt~\% solutions and for pure water (dot-dashed line) at $T$ = 293 K : (a) using criterion I, (b) using criterion II.
}
\end{figure}

Fig.~\ref{fig6} shows the populations $f_{j}$ of water molecules that form $j$ HBs with
their neighboring water molecules for the 66~wt~\% solutions and for pure water at $T$ = 293 K, using the two criteria previously defined.
Distributions using criterion II show a shift to larger numbers of water-water HBs compared to criterion
I because it is less stringent. As expected,
the maximum of the distribution for pure water is 4 for criterion II. Nevertheless, the distribution
is quite broad as a consequence of thermal disorder and may exceed 4 owing to the bifurcated HBs which are likely to occur.
Addition of disaccharide molecules to the solutions leads to a significant shift of the maximum of the
distribution toward lower values because of the formation of water-disaccharide HBs.
Trehalose appears to increase populations of water molecules with low $j$ (below 2 and 3 with (I) and (II), respectively) and to decrease
populations of water molecules with high $j$ (above 2 and 3 with (I) and (II), respectively) more than sucrose and maltose. This result is a
direct consequence of the difference observed on the hydration numbers.
Therefore, the tetrahedral HBN of water will be more disrupted in presence of trehalose than with
sucrose or maltose. This agrees with numerous experimental studies showing a higher \emph{destructuring
effect} of trehalose compared to sucrose and maltose as it has been particularly
demonstrated by Magaz\`u \emph{et al.}~\cite{Branca1999b}
from Raman scattering experiments.

\begin{figure}
\includegraphics[width=8.2cm,clip=true]{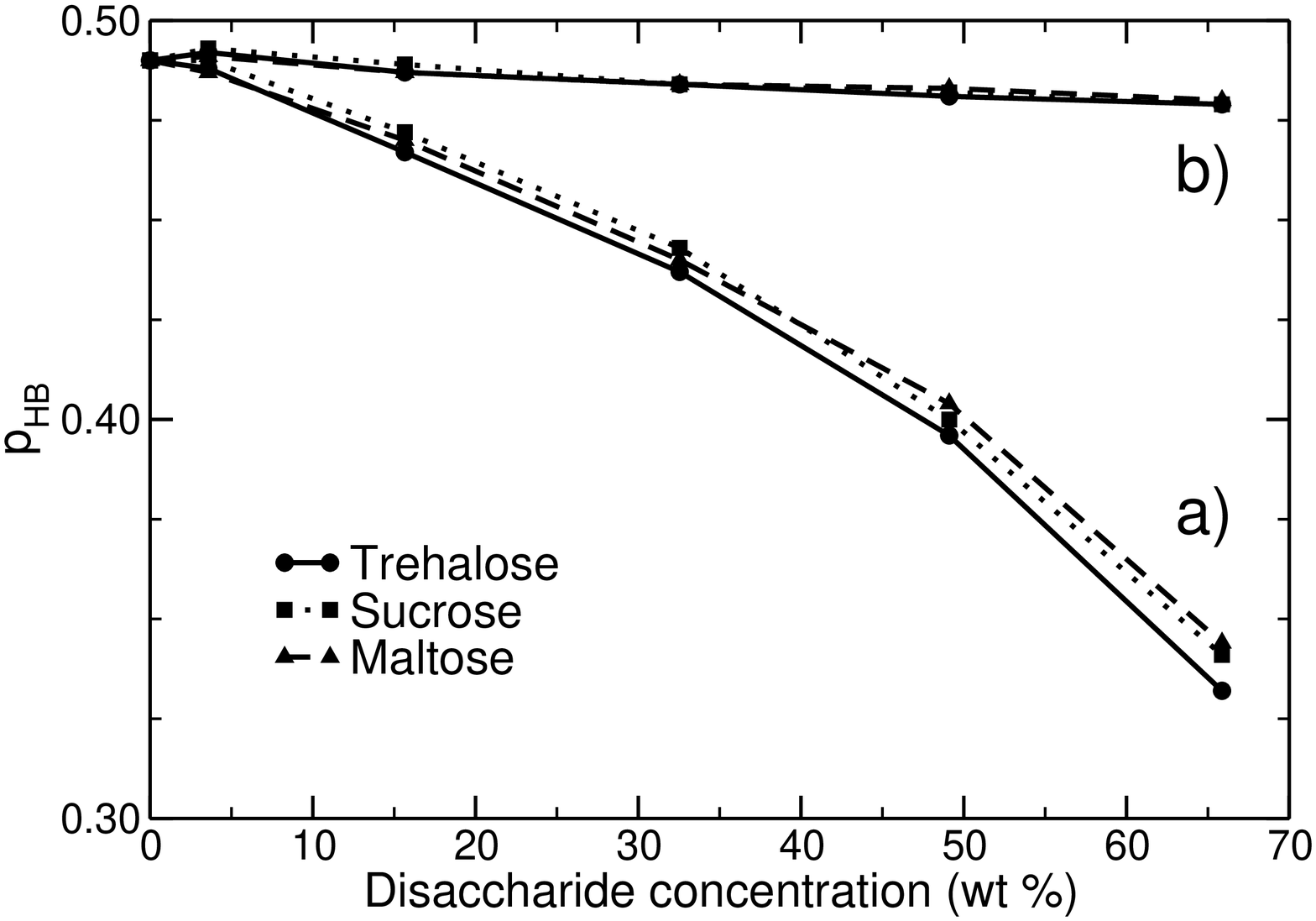}
\caption{\label{fig7}
Concentration dependence of the probability of HB formation $p_{HB}$ as deduced from Eq.~\ref{fj}
for the disaccharide solutions at $T$ = 293 K using criterion I : (a) for all water molecules
(b) excluding water molecules of the first hydration shell (defined using criterion II).
}
\end{figure}

By referring to the concept of intact and broken bonds, it is possible to deduce $p_{HB}$, the probability
of forming an intact bond from the previous distributions. Indeed, assuming that the HB formation is not
cooperative~\cite{Luzar1996} and the coordination number
in liquid water is 4, the fraction $f_{j}$ follows the binomial distribution~\cite{Stanley1980} :
\begin{equation}
f_{j} = \binom{4}{j}.p^{j}_{HB}.(1-p_{HB})^{4-j}\hbox{\hspace{0.1cm}.} \label{fj}
\end{equation}
Fig.~\ref{fig7} shows the concentration dependence of $p_{HB}$ of the disaccharide solutions at 293 K obtained from Eq.~\ref{fj} using the geometric
criterion I, which was chosen so that $j$ ranges from 0 to 4 for pure water ($f_{5}$ is negligible with criterion I).
Attempts to compute $p_{HB}$ with criterion II did not lead to satisfactory fits of Eq.~\ref{fj} using 5 or 6 
as the maximum coordination number
  and thus
are not presented.
The observed decrease of $p_{HB}$ mainly arises from the substitution of water-water HBs by
water-sugar HBs when increasing disaccharide concentration. As a consequence, the evolution of $p_{HB}$ as function 
of the concentration
is directly controlled by the sugar hydration numbers discussed previously. 
This is well demonstrated in Fig.~\ref{fig7}~b) when water molecules from the first shell were excluded.
Indeed, $p_{HB}$ clearly becomes
less dependent on the sugar
concentration and nearly constant. This figure also shows 
that the difference between sugars is 
largely reduced beyond the first-neighbor 
water layer. When the first hydration shell is excluded, 
no difference between sugars is observed.
This conclusion was already highlighted in our preliminary numerical work~\cite{Bordat2004} from the distributions of the orientation
parameter $q$ defined by Debenedetti \emph{et al.}~\cite{Debenedetti2001}. As a consequence, it appears that hydration numbers mainly
 rule
the difference observed among sugars, within the investigated concentration range. 
The higher hydration number of trehalose explains its
enhanced ability to disturb the HBN of water, revealed by the lowest $p_{HB}$ for trehalose-water solutions.
It should be noted that these results are in fair agreement with those obtained experimentally by
Branca \emph{et al.}~\cite{Branca1999b}.

\subsection{Water clusters}

Finally, we consider water clusters formed in the disaccharide-water solutions. Clusters of class $i$ were defined as ensembles of water
molecules H-bonded with each other and forming at least $i$ water-water HBs according to criterion I. Criterion I is more
suited to probe local tetrahedral arrangement of water molecules since it characterizes well-formed HBs, on the contrary to criterion II which
allows distorted configurations of water molecules. Alternatively, clusters could be defined using an order parameter such as the orientation
parameter $q$ suggested by Debenedetti \emph{et al.}~\cite{Debenedetti2001}. In this case, a cluster is defined as an ensemble of water molecules which have a $q$ parameter greater than a threshold value $q_{c}$~\cite{Errington2002} and are separated by a distance smaller than $r_{c}$.

\begin{figure}
\includegraphics[width=8.2cm,clip=true]{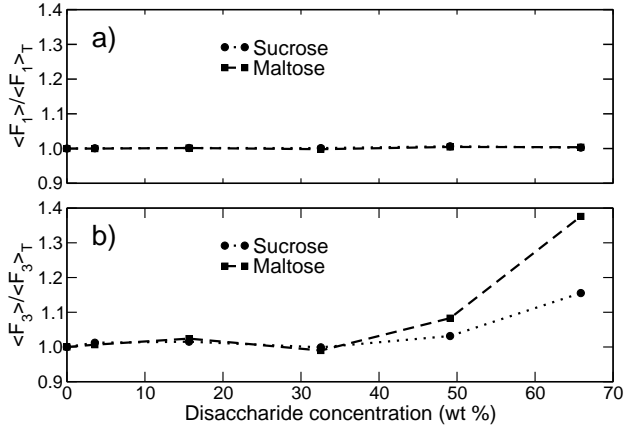}
\caption{\label{fig8}
Concentration dependence of the normalized ratios (a) $<F_{1}>/<F_{1}>$$\mathrm{_{T}}$ and (b) $<F_{3}>/<F_{3}>$$\mathrm{_{T}}$ of sucrose 
and maltose solutions at $T$ = 273 K, using criterion I. $<F_{i}>$ stands for the time-averaged fraction of water molecules involved in 
clusters of class \emph{i} and the subscript T refers to the trehalose solutions.
}
\end{figure}

On one hand, clusters of class 1 encompass both water molecules in a tetrahedral arrangement and in more distorted configurations.
Ambient water may then be described as a space-filling percolating HBN, as shown in previous papers~\cite{Geiger1979,Bordat2004}.
On the other hand, clusters of class 3 or 4 involve only water molecules in nearly tetrahedral configurations. Such clusters are thus
relatively small at ambient temperature and may be viewed as precursor nuclei of crystallization of ice.
The addition of sugars makes the probability of finding large numbers of such clusters very small. Therefore, we focus our attention
on the analysis of the proportion of water molecules included in each class $i$
as a means to clearly distinguish the effects of the three disaccharides. We define $<F_{i}>$ the fraction of water
molecules involved in clusters of class \emph{i} ($< >$ means time-averaging over the simulation length).
Fig.~\ref{fig8} shows the fraction of water molecules
$<F_{i}>$ of sucrose and maltose for clusters of class 1 and 3 at $T$ = 273 K normalized
by $<F_{i}>$$\mathrm{_{T}}$, the fraction of water molecules forming clusters of class $i$ in presence of trehalose.
$<F_{1}>$ of trehalose, maltose or sucrose are found identical at all concentrations.
This indicates that the total number of water molecules involved in clusters
is nearly independent on the studied sugars whatever the considered concentration, at $T$ = 273 K. In other words, the fraction of
water molecules not H-bonded with at least one water neighbor is essentially the same in the various disaccharide solutions.
However, water molecules reorganize in clusters of different sizes in the trehalose, sucrose and maltose solutions at concentrations above
$\phi_{A}\approx$ 40~wt~\%~\cite{Bordat2004}.
Indeed, the mean cluster size $<n_{W}>$ of water clusters at these concentrations is the lowest in trehalose solutions,
trehalose breaking clusters into smaller ones more efficiently (see Fig. 1 of~\cite{Bordat2004}), well in line with its higher hydration number.
As shown in Fig.\ref{fig8}b, the $<F_{3}>/<F_{3}>$$\mathrm{_{T}}$ ratios of sucrose and maltose exhibit a significant departure from 1.0
at concentrations above $\phi_{A}$. Most of the water molecules feel the presence of
disaccharides above $\phi_{A}$, owing to comparable numbers of water-sugar and water-water HBs. That is the reason why discrepancies
among the three disaccharides raise and why trehalose is found to reduce the number of locally tetrahedral water regions
- precursors of ice crystallization - more than sucrose and maltose (see also Fig.~\ref{fig7}). These ice nucleators become small and rare at these high
concentrations, at $T$ = 273 K. For example, only 6 to 9~\% of water molecules participate in clusters of classes 3 and 4 in the
66~wt~\% solutions at $T$ = 273 K. This may be understood as the very low probability of finding a water molecule surrounded by four other
neighboring water molecules not forming too many HBs with sugars, so that they can adopt positions and orientations compatible with a regular H-bonded
water tetrahedron.
Moreover, sucrose is found to disturb the tetrahedral water structure more than maltose, as discussed above for
hydration numbers. Consequently, hydration numbers of disaccharides play a key role at these concentrations.

\section{Sugar intramolecular and intermolecular interactions}

\subsection{Molecular flexibility}
\begin{figure}
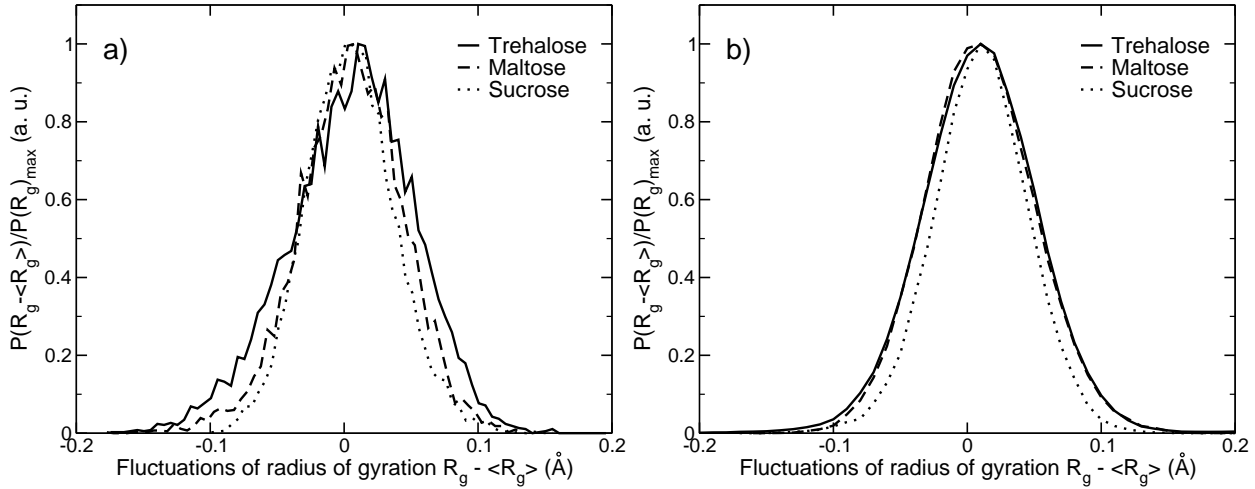

\includegraphics[width=8.2cm,clip=true]{figure8.eps}
\includegraphics[width=8.2cm,clip=true]{figure8_bis.eps}
\caption{\label{fig3}
Rescaled distributions of fluctuations of the radius of gyration $R_{g}$ of trehalose, sucrose, and maltose
at $T$ = 293 K for (a) the 4 wt~\% solutions and (b) the 49 wt~\% solutions. $<R_{g}>$ denotes the time-averaged 
radius of gyration and $P(R_{g})_{max}$ stands for the maximum of probability of the distribution $P(R_{g})$.}
\end{figure}

The intrinsic size of disaccharides can be roughly estimated from
their radius of gyration $R_{g}$, defined as
$R_{g}^2 =\frac{\sum{(m_{i}.r_{i}^2)}}{\sum{m_{i}}}$, where $i$ denotes the index number
of atoms within a given molecule, $m_{i}$ and $r_{i}$ the related mass and distance from the
center of mass of the molecule.
Fig.~\ref{fig3} show the disaccharide radius of gyration rescaled distributions $P(R_{g}-<R_{g}>)/P(R_{g})_{max}$
for the 4 wt~\% solutions at $T$ = 293 K.
Fluctuations of $R_{g}$ for trehalose molecules are clearly seen broader and more asymmetric than for
maltose or sucrose. This higher asymmetry may reflect the greater anharmonicity of the region of the PEL
explored by trehalose and is consistent with the larger amplitude of ring motions. Indeed, the number of intramolecular HBs in trehalose
is fewer compared to maltose and sucrose. These HBs restrict movements of disaccharides rings and of hydroxyl groups involved in HBs, thus
confining sucrose and, to a lower extent, maltose into more harmonic regions of the PEL \emph{i.~e.} closer to the local minimum.
Consequently, trehalose seems to be more flexible than maltose and sucrose, in the
4 wt~\% solution at $T$ = 293 K.
Most of this flexibility arises from small variations of the torsional angles about the glycosidic
linkage bonds, since rings are relatively rigid and stabilized by high energy barriers\cite{Liu1997}.
It should also be mentioned that the radius of gyration $R_{g}$ corresponding to
the maxima of the $P(R_{g})$ distributions are equal to 3.40~\AA \ , 3.45~\AA \ and 3.13~\AA \
for trehalose, maltose and sucrose, respectively. The lower radius
of sucrose comes from its more spherical structure compared to maltose and trehalose. Moreover, it contains a glucose ring and a smaller
fructose ring, whereas maltose and trehalose possess two glucose rings. The lower $R_{g}$ of trehalose
stems from the higher degree of symmetry of glucose rings relative to the glycosidic bond. Indeed,
glucose rings are linked through 
$C_{1}-O_{1}$ and $O_{1}-C_{1}'$ bonds for trehalose, and $C_{1}-O_{1}$ and $O_{1}-C_{4}'$ bonds for maltose (see Fig.~\ref{fig0}).

\begin{figure}
\includegraphics[width=8.2cm,clip=true]{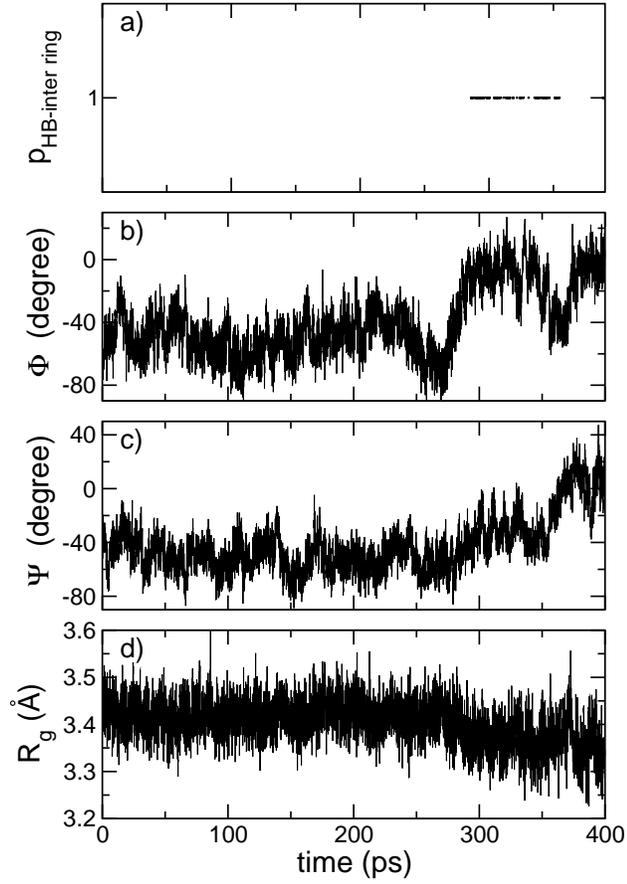}
\caption{\label{fig5}
Time-dependence of different parameters obtained for trehalose
in the 4 wt~\% solution at $T$ = 293 K:
(a) Probability of formation of an inter-ring HB $p_{HB-inter~ring}$, (b), (c) $\Phi$ and $\Psi$ glycosidic torsion angles 
defined as the dihedral angles between atoms $\mathrm{H_{1}-C_{1}-O_{1}-C_{1}'}$ and
$\mathrm{C_{1}-O_{1}-C_{1}'-H_{1}'}$ respectively (see Fig.~\ref{fig0}), 
and (d) radius
of gyration $R_{g}$.}
\end{figure}

In Fig.~\ref{fig5}, fluctuations of the glycosidic dihedral angles and the gyration radius $R_{g}$ are compared for trehalose.  
It should be noted that fluctuations are larger for trehalose than for maltose and sucrose, in good agreement
with the $R_{g}$ distributions discussed previously. ($\Delta\Phi,\Delta\Psi$) within the main minimum 
is ($13^{\circ}$,$12^{\circ}$), ($9^{\circ}$,$10^{\circ}$), ($11^{\circ}$,$7^{\circ}$)  and 
($11^{\circ}$,$13^{\circ}$), ($9^{\circ}$,$9^{\circ}$), ($9^{\circ}$,$10^{\circ}$)
for the trehalose, maltose and sucrose 4 and 49~wt~\% solutions, respectively - 
$\Phi$ and $\Psi$ are defined as the dihedral angles between atoms $\mathrm{H_{1}-C_{1}-O_{1}-C_{1}'}$ and
$\mathrm{C_{1}-O_{1}-C_{1}'-H_{1}'}$ for trehalose, $\mathrm{H_{1}-C_{1}-O_{1}-C_{4}'}$ and $\mathrm{C_{1}-O_{1}-C_{4}'-H_{4}'}$
for maltose, and $\mathrm{O_{5g}-C_{1g}-O_{1g}-C_{2f}}$ and $\mathrm{C_{1g}-O_{1g}-C_{2f}-O_{5f}}$
for sucrose (see Fig.~\ref{fig0}). This is indicative of a greater
conformational flexibility and suggests that the free energy minimum that trehalose
explores during our simulations is broader than that of maltose and sucrose.
This is also in agreement with the number of intra-HBs, which is the lowest
for trehalose and the highest for sucrose. Within a given well, intra-HBs prevent large motions
of rings.
This is illustrated over the 400 ps simulation of a 4 wt~\% trehalose
solution at $T$ = 293 K in Fig.~\ref{fig5}. During the first 260 ps of the simulation,
no HB forms between the rings of trehalose (see Fig.~\ref{fig5}a). Meanwhile, the glycosidic dihedral angles
$\Phi$ and $\Psi$ and the radius of gyration $R_{g}$ fluctuate around rather stable values~: $\Phi = -50~\pm~13^{\circ},
\Psi = -51~\pm~12^{\circ}, R_{g} = 3.42~\pm~0.04$~\AA~(see Fig.~\ref{fig5}b, c, d, respectively).
After 260 ps, the dihedral angle $\Phi$ shows a
transition to $\Phi$ = -14~$\pm$~16$^{\circ}$
while for $\Psi$ the transition is less pronounced until $t$ = 350~ps ($\Psi$ = -22~$\pm$~22$^{\circ}$).
This transition is connected with the formation of an intramolecular HB between $\mathrm{O_{2}}$ and $\mathrm{O_{6}'}$ oxygen atoms. This intra-HB
was observed by Conrad \emph{et al.} at high concentrations (50 wt~\% or more) at 358~K~\cite{Conrad1999}.
As a direct consequence of this conformational transition,
the radius of gyration $R_{g}$ of trehalose is reduced - $R_{g} = 3.37~\pm~0.05$~\AA \ - suggesting a more compact conformation,
rings becoming closer to each other. We also expect smaller fluctuations of $R_{g}$. However, the short time during which this new conformation
is observed does not allow us to confirm.

Besides, the intrinsic enhanced flexibility of trehalose is not contradictory with its lower mean-square fluctuations $<u^2>$ compared
to sucrose, as experimentally observed~\cite{Magazu2004}. 
Indeed, the motion of hydrogen atoms both depend on the global molecular translation and rotation and
on the internal vibrations. Therefore, trehalose may show a greater flexibility (probed \emph{e.g.} by the radius of gyration fluctuations)
and a smaller $<u^2>$ compared to sucrose. This point will be described in more details in a next paper devoted to the dynamical
features of sugar-water solutions~\cite{Affouard2005}.


Fig.~\ref{fig5} underlines the close relation between intramolecular HBs and molecular conformational
fluctuations. In particular, it may confirm
the hypothesis of Oku~\emph{et al.}~\cite{Oku2003} who investigated the mechanism of the antioxidant
function of trehalose by means of NMR and quantum chemistry.
They proposed that trehalose interacts specifically with one \emph{cis} double bond of an unsaturated
fatty acid via the $\mathrm{O_{6}-H_{6}'}$ and either $\mathrm{O_{2}-H_{2}}$ or $\mathrm{O_{3}-H_{3}}$
 hydroxyl groups of trehalose. The complex formed was also stabilized by O-H$\cdot\cdot\cdot$$\pi$ and C=H$\cdot\cdot\cdot$O types of HBs.
The greater conformational flexibility we observe for trehalose in the present study
might favor these kinds of interactions compared to sucrose and maltose. Indeed, it may allow the hydroxyl groups
mentioned above to adopt the required conformation, while trehalose remains in the \emph{gauche}
conformation before and after complexation \emph{i. e.} in the same energy well. Furthermore, the occurrence of an
inter-ring HB between $\mathrm{O_{2}}$ and $\mathrm{O_{6}'}$ oxygen atoms of trehalose observed in the 4 wt \% solution
may indicate that the stable conformation of trehalose is favorable to form such a complex. On the contrary, the
more frequent inter-ring HBs found in maltose and sucrose as well as their intrinsic topology might prevent
such interactions.
Accordingly, the enhanced conformational flexibility of trehalose may favor HBs
with polar groups of membranes or proteins as proposed by
Crowe \emph{et al.}~\cite{Crowe1984} and referred to the \emph{water replacement} hypothesis.

\subsection{Sugar clusters}
\label{sugar_cluster}
As the sugar content increases in the studied solutions, sugar-sugar interactions become more and more likely (see Table~\ref{table3}).
At a given concentration threshold, sugar clusters may even percolate, forming a continuous space-filling network.
The percolation of the sugar HBN may explain the decoupling of water and sugar diffusions. Indeed, it has been
observed experimentally that sucrose and water diffusions decouple at water contents below $\approx$~50~wt~\% ~\cite{Rampp2000}.
Moreover, Goddard~\emph{et al.}~\cite{Molinero2003} have suggested from simulations that the formation of sucrose network in highly
concentrated solutions may increase both the resistance to shear deformation and the mechanical stability of the mixture.

The formation of a sugar network should be favored by a low number of intra-HBs, since this kind of HBs involves OH groups that
are not available for inter-HBs. Moreover, the number of sugar-sugar inter-HBs should reveal clustering tendencies.
The largest number of intermolecular sugar-sugar HBs found for maltose
(see statistics given in Table~\ref{table3}) may imply that the molecules of this sugar
 tend to organize themselves into
larger clusters than for sucrose or trehalose. In order to describe the sugar cluster formation suggested by these HB data, we define a cluster
as a set of sugar molecules connected to each other by at least one HB according to criterion II. This geometric criterion
was preferred to criterion I in order to avoid scarce clusters and consequently very poor statistics.
The mean sugar cluster size $<n_S>$ was computed as
$<n_{S}> = \sum{n_{S} . W_{S}}$, where $n_{S}$ is the number of sugars composing a given cluster and $W_{S}$ is
the probability of the clusters of size $n_{S}$. Isolated sugar molecules are considered as clusters of size one.
Since sugar dynamics are much slower than water dynamics~\cite{Feick2003} and
their number is relatively small in our simulation boxes (between 1 and 52), the statistics of cluster formation is rather poor for a given
simulation. Therefore, since no significant differences of $<n_{S}>$ as function of the temperature were observed, we have
averaged $<n_{S}>$ over the (293~K-373~K) temperature range at a given concentration (the results at $T$ = 273 K have not been used
because of the lack of sugar diffusion at high concentrations). Four aberrant values over 60
were discarded (2 for sucrose and 1 for maltose and trehalose, among the 4 concentrations and 5 temperatures considered for these calculations).
Consequently, $<n_{S}>$  provide a rough trend of the sugar cluster formation as a function of concentration.

\begin{figure}
\includegraphics[width=8.2cm,clip=true]{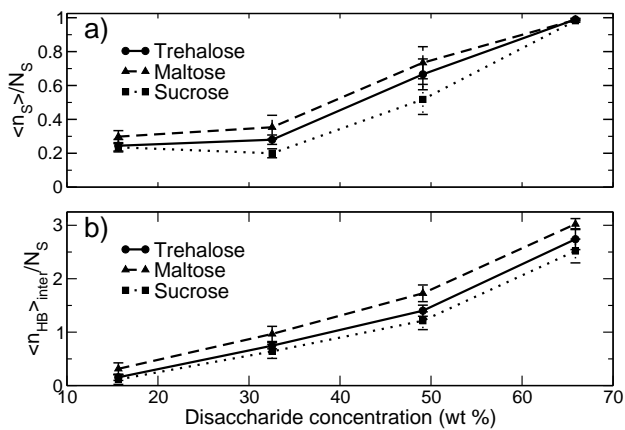}
\caption{\label{fig9}
Concentration dependence of (a) the normalized mean sugar cluster size $<n_{S}>$/$N_S$ and (b) of
the normalized mean number of sugar-sugar HBs $<n_{HB}>_{inter}$/$N_S$ of the disaccharide solutions, averaged
over the (293~K-373~K) temperature range to improve statistics (Table~\ref{table3} gives $<n_{HB}>_{inter}$/$N_S$ at $T$ = 293 K). 
Error bars are calculated from the standard
deviations. $<n_{S}>$/$N_S$ and $<n_{HB}>_{inter}$/$N_S$ cannot be calculated for our most diluted
solutions (4~wt~\%), because there is only one sugar in the simulation box.
}
\end{figure}

Fig.~\ref{fig9}a shows the evolution of the ratio $<n_{S}>/N_S$, where $N_S$ is the total number of sugars in the simulation boxes, for the three
disaccharide solutions as a function of concentration. At low concentrations (16 and 33 wt \%) small
clusters may appear in the solutions, $<n_{S}>$/$N_S$ remaining roughly constant around 0.2~-~0.3.
Therefore, the mean size of cluster $<n_{S}>$ is rather proportional to the total number of sugars $N_S$. This behavior changes radically at higher
disaccharide concentrations, where sugar clusters grow more than linearly with $N_S$ and $<n_{S}>$/$N_S$ increases to one
for the solutions at 66 wt~\%, which can be considered as the treshold concentration of percolation, $\phi_p$. Accordingly, the percolation
of the sugar HBN has been achieved at this concentration \emph{i. e.} most of sugars of the
simulation box belong to the same large cluster. Notwithstanding the quite large error bars, differences below $\phi_p$ between maltose or
trehalose on the one hand and sucrose on the other hand emerge. Indeed, sucrose is found to form clusters of smaller sizes than trehalose and
maltose. This is probably related to the higher number of intra-HBs of sucrose
molecules compared to maltose or trehalose, since hydroxyl groups involved in these intramolecular interactions do not remain
available for other intermolecular interactions (with water or sugar molecules). As a consequence, a higher sucrose
concentration is needed to reach a given cluster size compared to maltose or trehalose.
Furthermore, $<n_{S}>$ of maltose and trehalose are found relatively close to each other, consistent with their very similar topology.
Nevertheless, maltose molecules have a propensity to form a larger number of HBs with each other than trehalose molecules. This is
illustrated in Fig.~\ref{fig9}b, which presents the normalized averaged number of intermolecular sugar-sugar HBs $<n_{HB}>_{inter}$/$N_S$ 
(see also Table~\ref{table3}). At low concentrations, this number is limited by the quite low probability
of forming inter-sugar HBs, which prevents multiple HBs between two given sugars. It increases with sugar concentration,
but is restricted at high concentrations by the maximum number of inter-HBs that sugars may form owing to sterical hindrance.
The averaged number of intermolecular sugar-sugar HBs is found larger 
in the maltose clusters than in the trehalose ones. Thus, maltose favors more sugar-sugar HBs
than trehalose and sucrose. This explains in part the lower hydration number of maltose relative to trehalose at high
concentrations since sugar-sugar HBs reduce the possibility that the water molecules have to bind to sugars.

\begin{figure} 
\includegraphics[width=7.0cm,clip=true]{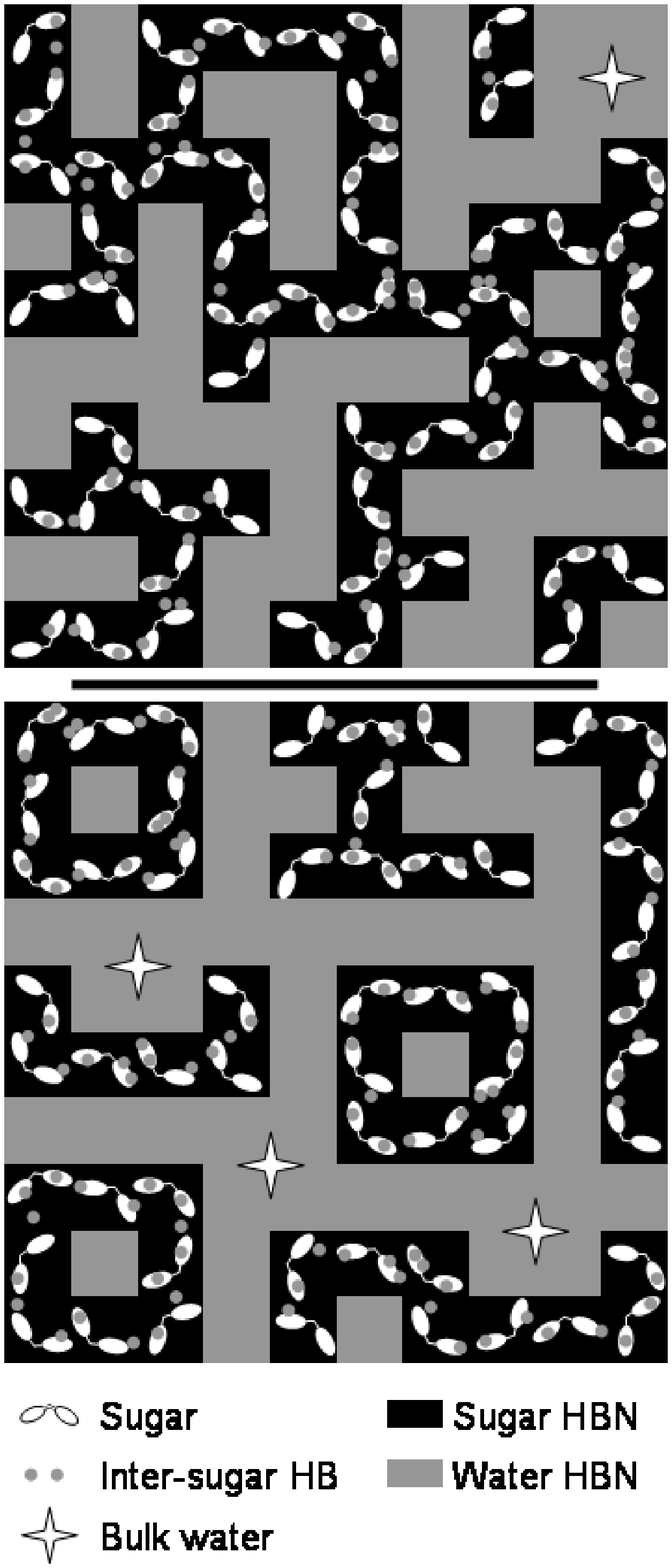}
\caption{\label{fig10}
\baselineskip=.5cm
Schematic 2D-representation of a microscopic region of trehalose (top) and sucrose (bottom) solutions in the intermediate concentration
range (33~wt~\% - 66~wt~\%) \emph{i.~e.} below the percolation treshold $\phi_p$. The differences between these two cartoons are purposefully
magnified in order to clearly distinguish the structural features of trehalose and sucrose solutions.
Grey parts represent water molecules (whether they are H-bonded to sugars or not), while black parts stand for the sugar molecules.
Stars designate sites where \emph{bulk water} - not in the first hydration shell of sugars - may exist.
Trehalose molecules form quite large and interconnected clusters, whereas sucrose arranges itself into smaller, isolated clusters. This makes
trehalose molecules less mobile than sucrose ones. Moreover, owing to the larger number of water-trehalose HBs, water molecules diffuse less in
trehalose-water mixtures, and sites where water may crystallize (\emph{bulk-like}) are more scarce than in sucrose solutions. This makes
water-trehalose mixtures more homogeneous, from a structural point of view at least.
}
\end{figure}

This peculiar interplay between the capability of a sugar to hydrogen-bond to other sugars or to water seems
critical for the homogeneity of the system and could probably play a significant role in the
preservation of bio\-mo\-le\-cu\-les.
Indeed, it suggests that below the percolation concentration $\phi_p$ trehalose
may form more homogeneous matrices with water with respect to sucrose or maltose. 
Trehalose would be able to form
large clusters with itself while maintaining higher
hydration numbers than sucrose or maltose. 
Moreover, since trehalose forms HBs with a
larger number of water molecules than sucrose and maltose do,
less \emph{bulk} water molecules (marked by a star in Fig.~\ref{fig10}) would exist in the trehalose solutions. Therefore,
both the ice formation probability and the dehydration stresses arising from the water escape would be reduced. However, this new insight suggests 
experiments to probe the influence of the sugar-water matrix nanostructure on the dehydration kinetics. 
Fig.~\ref{fig10} outlines schematic 2D-views of concentrated trehalose and sucrose aqueous solutions to illustrate this hypothesis. These representations
are based on the interpretation of the statistical calculations shown in Fig.~\ref{fig9} and therefore are difficult to observe for an instantaneous
configuration of our simulations trajectories.
It should also be reminded that sugar clusters reorganize continuously until percolation has been achieved.

\section{Summary and conclusion}
Structural properties of trehalose, maltose and sucrose disaccharides
in water have been investigated by
MD computer simulations. 

Crystallinity of sugar/water solutions have been studied from calculations of partial static 
structure factor found 
in good agreement with neutron diffraction experiments and 2D-radial distribution functions. 
A significant increase of the localization
of water molecules at the hydration sites of the trehalose dihydrate crystal is observed. It strongly supports
the more crystalline character of trehalose solution suggested in~\cite{Magazu2004}.
Hydration numbers have been accurately determined for
two different models of water (SPC/E and TIP3P - for the 4 wt \% solutions)
and different geometrical criteria aiming to distinguish both well-formed and less-formed water
HBs. We have clearly shown that trehalose binds to a larger number of water molecules than do maltose and sucrose, for all studied concentrations.
Our data have also been successfully confronted to experimental and other MD simulation results
of hydration numbers.
The structure of the water HBN in presence of sugars has been studied using
the concept of broken HBs proposed by Teixeira and Stanley~\cite{Stanley1980}.
Our results prove that the presence of disaccharide molecules in the investigated solutions leads to
a significant decrease of water populations with high coordination numbers and to
a smaller probability $p_{HB}$
of forming an intact bond, this effect being more salient in trehalose solutions.
However, the evolution of $p_{HB}$ as function
of the concentration
was also found to be directly controlled by the sugar hydration numbers. The discrepancy between 
the different sugars is largely reduced beyond the first hydration shell.
Statistics of clusters size made of hydrogen-bonded waters have been investigated and reveal that 
above a concentration $\phi_{A} \simeq 40 $ wt \%, trehalose is able to reduce the water cluster
size more efficiently than sucrose and maltose. 
This confirms the superior capability of trehalose to disturb the HBN of water and is consistent with
the \emph{destructuring effect} model suggested by Magaz\`u~\emph{et al.}~\cite{Branca1999b}.

From the fluctuations of the radius of gyration and of the glycosidic angles at high dilution, we have also shown that trehalose
exhibits a higher flexibility with respect to maltose and sucrose, consistent with its lower number of intramolecular HBs and thus
with its higher hydration number. This enhanced flexibility would confirm the hypothesis of Oku~\emph{et al.}~\cite{Oku2003} suggesting that
trehalose is able to form HBs with non-polar residues of proteins. 
Sugar-sugar interactions have been analyzed in terms of sugar clusters. Sugar molecules form a percolating HBN in the
66~wt~\% solutions. At intermediate concentrations, the averaged size of clusters $<n_{S}>$ of trehalose and maltose
is larger than $<n_{S}>$ of sucrose owing to fewer intra-HBs. Moreover, trehalose favors to a lower extent inter-sugar HBs
compared to maltose, thus preserving more efficiently HBs with water molecules.  The trehalose/water solution is found more homogeneous than
the others.
This particular property should play a major role in
the biopreservation efficiency of water-trehalose mixtures against desiccation stresses
and ice formation.

To conclude, the results reported here show that 
maltose/water systems are less homogeneous owing the tendency of maltose molecules to form clusters 
and thus reducing their possibility to destructure the water HB network. For sucrose/water solutions,
the higher probability of sucrose molecules to form intra-molecular HBs strongly reduces their interaction
with both water or other sugar molecules. Unlikely to maltose or sucrose, trehalose
molecules possess similar capabilities to interact with both sugar and water molecules which 
make trehalose/water a more homogeneous system.
This homogeneity of the disaccharide/water matrices due to the reciprocal interactions between sugar and water molecules
may rule their biopreservative efficiency. 
It may also be of utmost importance in the interactions between sugars and biomolecules. 
Lysozyme/disaccharides/water simulations are currently led to shed more light on disaccharide-protein interactions.
Results obtained in the present paper also call for new investigations to clarify 
the influence of the solution structural homogeneity on their dynamical properties~\cite{Affouard2005}.

\bigskip

{\bf Acknowledgments}

The authors are grateful to Prof.~A.~Ces\`aro for fruitful discussions and wish to acknowledge the use of the facilities of the IDRIS (Orsay, France) and
the CRI (Villeneuve d'Ascq, France) where calculations were carried out.
This work was supported by the INTERREG III (FEDER)
program (Nord-Pas de Calais/Kent).

\bibliographystyle{jpc}

\end{document}